\documentclass{osa-article}
\usepackage{cite}
\usepackage{amsmath}
\usepackage{braket}
\usepackage[]{graphicx}
\usepackage{bm}
\usepackage{booktabs}
\usepackage{makecell}
\usepackage{multirow}
\usepackage{color}
\usepackage[pagewise]{lineno}
\newlength{\figwidth} 
\newcommand{\beq}{\begin{equation}}
\usepackage{comment}
\usepackage{verbatim}

\newcommand{\beql}[1]{\begin{equation}\label{#1}}
\newcommand{\eeq}{\end{equation}}
\newcommand{\bsp}{\begin{split}}
\newcommand{\esp}{\end{split}}
\newcommand{\Eq}[1]{Eq.~(\ref{#1})}
\newcommand{\Equation}[1]{Equation~(\ref{#1})}

\newcommand{\Fig}[1]{Fig.~\ref{#1}}

\newcommand{\Table}[1]{Table.~(\ref{#1})}

\journal{oe}
\begin{document}

\title{Finite element based Green's function integral equation for modelling light scattering}
\author{ Wen Li,\authormark{1, $\dagger$}, Dong Tan,\authormark{1, $\dagger$}, Jing Xu, \authormark{1,2,3}, Shubo Wang\authormark{3}  and Yuntian Chen,\authormark{1,2}}
\address{\authormark{1}School of Optical Electronic Information,Huazhong University of Science and Technology,Wuhan,China \\
\authormark{2} Wuhan National Laboratory for Optoelectronics, Huazhong University of Science and Technology, Wuhan, China\\
\authormark{3}Department of Physics, City University of Hong Kong, Hong Kong, China\\
\authormark{$\dagger$}These two authors contributed equally to this work.}
\email{\authormark{3}yuntian@hust.edu.cn}   

\begin{abstract}
We revisit the volume Green's function integral equation for modelling light scattering with  discretization strategies as well as numerical integration recipes borrowed from finite element method. The merits of introducing finite element techniques into  Green's function integral equation are apparent. Firstly, the finite element discretization provides a much better geometric approximation of the scatters, compared with that of  the conventional discretization method using staircase  approximation. Secondly, the accuracy of  numerical integral inside one element associated with all different types of Green's function integral equations can be greatly improved by introducing  quadrature rules. Within the standard framework of Green's function integral equation, we seamlessly introduce finite element techniques into the Green's function integral equation by introducing the auxiliary variables that confines the singular integrand inside each element, leading to a better and more flexible approximation of the geometry of the scatters and a more accurate numerical integral. We then illustrate the advantages of our finite element based Green's function integral equation method  via a few concrete examples in modelling light scattering by optically large and complex scatters. 
\end{abstract}


\section{Introduction}
Green's function integral equation (GIE) is an elegant and efficient approach  of calculating the scattering of waves \cite{Tai1971,Economou}, and has been widely applied in modelling light scattering in nano-structures \cite{Lavrinenko,sondergaard2007modeling,Novotny,Chew2009,Martin,Jung2011dyadic,Chen2017general} and acoustic scattering \cite{Wilde}. An appealing advantage of GIE is that only the scatter is needed to be discretized, in contrast to other approaches, e.g., finite element method (FEM) and finite difference time-domain (FDTD) method, which require both the scatter and the background to be discretized. This leads to the reduction of the computational overhead, which can be further reduced by transforming the volume GIE \cite{Yurkin,Kottmann} into a surface GIE \cite{Gibson2015,Harrington,Morita,shubo,Jung2008greens,Waxenegger,Hohenester2012MNPBEM,Hohenester2014simulating,Prather,Myroshnychenko,Garcia,Kress}, where only the boundary of the scatter needs to be discretized. However, the challenge associated with volume GIE and surface GIE is the singular behavior of  integrand, which has attracted considerable attention in the last decade. Several schemes have been developed to address the challenge, include: singularity subtraction \cite{Hanninen},  improved numerical integration \cite{Chen2016accurate,Graglia,Polimeridis,Vipinaa}, and analytical approximation of the singular terms \cite{Nachamkin,Yaghjian,Sondergaard2018book}. 

Interestingly, considerable efforts have been dedicated to introducing FEM techniques \cite{Jin,Chen2010finite,Chen2016accurate}, e.g.,  advanced discretization strategies \cite{Sondergaard2018book,Kern} as well as efficient  algorithms  of carrying out numerical integrals \cite{Chew2009,Raziman}, into GIE. Indeed, the finite element discretization could not only yield an improved geometric approximation, but also improve the numerical accuracy of the integral by introducing the quadrature-rule based integration techniques that are widely used in FEM. For instance, triangle discretization has been introduced to approximate the two-dimensional (2D) scatters in volume GIE \cite{Sondergaard2018book}, or surfaces of 3D scatters in surface GIE \cite{Kern,Rodriguez,Chew2009}. Apart from the various advantages, the accuracy of GIE based on finite element discretization and quadrature integration rule is still compromised by the singular integrand, as studied  in Ref \cite{Raziman}, where the link between the numerical quadrature order and the integral accuracy is investigated in details. In particular, the authors in  Ref \cite{Raziman} find that the singular behavior of Green's tensor at the source causes a large variation in field amplitude over triangle elements close to it. Such large variation nearby the sources is due to the use of the divergence-conforming elements equipped with Rao-Wilton-Glison basis function \cite{Kern,Raziman,Chew2009},  which are defined on a pair of triangles sharing the same edge. 

In this paper, we elaborate the combination of  FEM techniques, i.e., triangle discretization and quadrature rule,  into GIE by introducing  auxiliary variables. In our proposed finite element based Green's function integral equation (FEGIE), we follow the same course of implementing GIE that is extensively studied by S\o ndergaard \cite{Sondergaard2007slow,Sondergaard2018book,Jung2009theoretical,Siahpoush,Sondergaard2012coupling,Jung2008greens,Sondergaard2013theoetical,Sondergaard2008strip}. In contrast to S\o ndergaard's work,  we use the auxiliary variables instead of the dependent variables to discretize GIE. The auxiliary variables are the field values with carefully selected coordinates, which coincide with the coordinates of  quadrature points inside each element. Given the dependent variables of the triangles and the associated basis functions, the auxiliary variables can be easily interpolated. The purpose of introducing the auxiliary variables is to confine the singular behavior of integrand inside its own triangle, such that the singular term can to a good approximation be integrated analytically one by one. The second benefit of introducing auxiliary variables is the improved integral contributions from other triangles, apart from the singular contribution from its own triangle. Lastly, the auxiliary variables  can be seamlessly and conveniently integrated into the current framework of GIE with an efficient assembly strategy. As an illustration of this idea, we use triangle elements to discretize the scatter, and provide two types of implementation strategy to realize FEGIE \cite{code}. 

The paper is organized as follows. In Section 2, we give a brief yet complete description of volume Green's function integral equation, accompanied with two different implementation strategies, i.e., type-1 FEGIE and type-2 FEGIE. In Section 3, we use FEGIE to calculate the far-field pattern of single circular rod, 3-layered rod, unidirectional 5-layered rod, and a  complex scatter with corrugated surface. We find that the two schemes in FEGIE yield considerably accurate far-field pattern  compared with that  obtained from Mie theory  \cite{schafer2011,schafer2012,lee,kerker,bohren2008} or commercial software package: Multiphysics COMSOL \cite{comsol}, and  yield more accurate far-field pattern compared with the standard discretization scheme, i.e., staircase approximated with square grids. Finally, Section 4 concludes the paper.

\section{Principles of finite element based  Green's function integral equation}
Without losing the generality, we illustrate the seamless integration of finite element techniques into the volume GIE with a better geometric approximation as well as smaller number of degrees of freedom (DOFs). Notably, the extension to other types of Green's function integral equation, i.e., surface GIE equation, is straightforward and can be implemented in a similar fashion. 
\subsection{Introduction of volume Green's function integral equation in the modelling of light scattering}

We give a brief introduction  to the volume GIE for completeness. Provided a current source with its  current  density   $\bm{J}_s (\bm r)$ distributed in the domain $\Omega_s$,  the electric field $\bm E (\bm r)$ in the whole computational domain $\Omega$ can be given by 
\beq\label{eq1}
\bm E\left( {\bm{r}} \right) =  - i\omega {\mu _0}\int _{\Omega_s} {\bar {\bm g}\left( {{\bm{r}},{\bm{r'}}} \right)} \bm{J_s}\left( {{\bm{r'}}} \right)dV, 
\eeq
where $\bar{\bm g}\left( {{\bm{r}},{\bm{r'}}} \right)$ is the dyadic Green's function determined  by $[\nabla  \times  \nabla  \times  - k_0^2 \bar{\bm {\varepsilon}}_r (\bm r)] \bar {\bm{g}} (\bm r,\bm r' )= \bar {\bm {I}}_0\delta(\bm r-\bm r')$.

Next, we consider a plane wave ($\bm E_0(\bm r)$) incident upon a scatter (dielectric function given by $\bar {\bm\epsilon}_s(\bm r)$) embedded in the hosting medium ($\bar {\bm\epsilon}_{bg}$), and we intend to compute  the scattering field $\bm E_s(\bm r)$ in $\Omega$. Importantly, both the incident field $\bm E_0$ and  the total field $\bm E_{tot}(\bm r)=\bm E_{0}(\bm r)+\bm E_{s}(\bm r)$ satisfy the following homogeneous wave equation, 
\begin{subequations}\label{SameBeta}
\begin{align}
[\nabla  \times  \nabla  \times  - k_0^2 \bar{\bm {\varepsilon}}_{bg} ]  \bm{E}_0(\bm r)=0, \label{background} \\ 
[\nabla  \times  \nabla  \times  - k_0^2 \bar{\bm {\varepsilon}}_r (\bm r)]  \bm{E}_{tot}(\bm r)=0. \label{scattering}
\end{align}
\end{subequations}
According to the linearity of vector wave equation, one can derive the relation between  total and incident field via the background Green's function,
\beq\label{volumegreen3D}
\bm E_{tot}\left( {\bm{r}} \right) = \bm {E}_0\left( {\bm{r}} \right) + \int_ {\bm{r'} \in \Omega}{\bar{\bm g}\left( {{\bm{r}},{\bm{r'}}} \right)} k_0^2\left( {\varepsilon \left( {{\bm{r'}}} \right) - {\varepsilon _{ref}}} \right)\bm E\left( {{\bm{r'}}} \right)dv.
\eeq
\Equation{volumegreen3D} is the volume GIE in the full vector form, describing the scattering of three-dimensional (3D) scatters. In order to illustrate the basic concepts of  how the FEM techniques can be useful for GIE, we consider a 2D scattering object under the illumination of TE polarization, in which \Eq{volumegreen3D} can be reduced into a scalar form,
\beq\label{volumegreen2D}
E_z\left( {\bm{r}} \right) =  {E}_0^z\left( {\bm{r}} \right) + \int_ {\bm{r'} \in \Omega}{g\left( {{\bm{r}},{\bm{r'}}} \right)} k_0^2\left( {\varepsilon \left( {{\bm{r'}}} \right) - {\varepsilon _{ref}}} \right)E_z\left( {{\bm{r'}}} \right)dv,
\eeq
where ${g\left( {{\bm{r}},{\bm{r'}}} \right)}$ is the 2D Green's function, whose explicit expression and its usage to calculate the far-field pattern given in Appendix A.  


\subsection{Finite element based Green's function integral equations}
In the following, we proceed to discuss the geometric discretization of the scatter as well as the numerical discretization of GIE. The standard procedure of implementing FEM calculations basically contains 4 steps \cite{Jin}: (1) using the grids or elements to approximate complex structures; (2) selecting proper  polynomial  functions with order $n$ to interpolate the field within an element; (3) using Galerkin method to formulate FEM problems defined by certain partial differential equations; (4) Assembling the matrix and solving the set of linear equations. Our proposed FEGIE method has the same 4 steps, except that the FEM formulation in step (3) is replaced by the formulation of the discretized GIE. In FEM, the dependent variables are discretized values of the field, the coordinates of which are carefully chosen according to  the grids and the interpolated functions. With the assistance of quadrature rules, the numerical integration in FEM can be highly accurate by interpolating the field values at quadrature points. In contrast, the convergence of numerical integration in the discretized GIE is more difficult to achieve than that in FEM, due to the existence of singular terms in the integrand of \Eq{volumegreen2D}. The singular terms are generated at $\bm r$ = $\bm r'$ in the Green's function. This is in conflict with the standard FEM implementation, since the dependent variables are defined at the vertices shared by a number of triangles, which causes the sharing of singular behavior of the integrand among neighbour triangles.

To overcome this difficulty,  the auxiliary variables,  i.e., the field values used in the  discretization  of GIE, is introduced to remove the singular behavior of integrand from the nearby triangles. Once the singular behavior is limited inside its own triangle, analytical approximation or singularity subtraction techniques can be applied to obtain a relatively accurate integration. The auxiliary variables can be interpolated from the dependent variable, but not necessarily the same as the dependent variables. This subtle difference between auxiliary variables and  dependent variables  that are overlooked in literature \cite{Sondergaard2018book,Kern,Raziman,Chew2009} leads to two different  strategies of implementing  FEGIE, i.e., named as type-1  FEGIE and type-2 FEGIE, which will be discussed in this paper. In type-1  FEGIE, auxiliary variables and  dependent variables  are indeed identical, which is easier to implement but contains redundant DOFs. In type-2 FEGIE, auxiliary variables and  dependent variable  are different, which leads to the significant reduction of DOFs and conceptual difficulty that will be addressed shortly.

\begin{figure}[ht!]
\centering
\includegraphics[width=12cm]{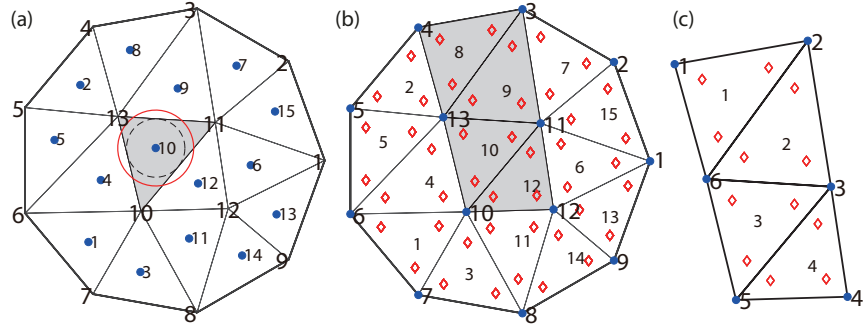}
\caption{A circular disk is discretized into 15 triangles which are further used in (a)  type-1 FEGIE and (b)  type-2 FEGIE.  (c) A small portion, indicated by light gray, of the circular disk of (b) with only 4 triangles and 6 vertices. In  type-2 FEGIE, the dependent variables are the vertices indicated by the solid blue dots, while the auxiliary variables associated with each triangle are the quadrature points indicated by red diamonds inside the corresponding triangles.}
\label{fig1}
\end{figure}
\subsubsection{Type-1 FEGIE}
We first consider a simple scenario, i.e., type-1 FEGIE. As an example shown in \Fig{fig1}(a), the locations of  dependent variables and  auxiliary variables in type-1 FEGIE coincide at the origins, i.e., indicated by the blue dots, of the in-circles of 15 triangles. With discretized triangles labelled by $i$, one can reformulate \Eq{volumegreen2D}  into  discretized GIEgiven by 
\beq\label{discre_VolumeGreen2D}
{E_i} = {E_{0,i}} + \sum\limits_j {{g_{ij}}k_0^2\left( {{\varepsilon _j} - {\varepsilon _{ref}}} \right)} {E_j}{A_j}
\eeq
where  $A_j$ is the area of the $j$th triangle. As $i=j$, the integrand in \Eq{volumegreen2D}  becomes singular and can be calculated by approximating the triangle by a circle with the same area. Taking the  gray triangle in \Fig{fig1} (a) for example, the equivalent circle is shown as the red circle, which has the radius given by $r=\sqrt{\frac{A_{10}}{\pi}}$. This treatment leads to a relatively accurate approximation of $g_{ii}$ by analytically integrating the singular integrand of \Eq{volumegreen2D}  within the red circle. Once the descretized GIE, i.e., \Eq{discre_VolumeGreen2D}, is known, it is straight forward to calculate the dependent variable $E_i$, and thus the field values at all the other coordinates (see explicit expression $g_{ii}$  and the far-field pattern calculation in appendix A).

\subsubsection{Type-2 FEGIE}
In contrast to type-1 FEGIE, we consider the second scenario where the dependent variables and the auxiliary variables are different, as illustrated in \Fig{fig1} (b). As shown in \Fig{fig1} (b), the dependent variable are field values at the vertices, while auxiliary variables that are used to discretize GIE are the field values at the red diamonds. Though the triangulation in \Fig{fig1} (a) and \Fig{fig1} (b) is identical for the same circular disk, the different treatments of the dependent variables and auxiliary variables lead to completely different implementations of FEGIE, which are the main concerns of this paper. To further illustrate our strategy to implement the  discretized GIE, i.e., the aforementioned step (3) similar to FEM modelling,  we consider the simplest discretization of a scatter (containing only  6 vertices and 4 triangles),  as shown in \Fig{fig1} (c), and show how to construct type-2 FEGIE through this simple example with first order quadrature rule and first order interpolating function.

Firstly, we use six dependent variables, i.e., the field values at the  6 vertices, to interpolate field values at all the other locations, including the  auxiliary variables with the locations indicated by the red diamonds. For instance, as shown in \Fig{fig1} (c), the field value $u^i(x,y)$  at position $(x,y)$ inside the $i$th triangle  is given by $u^i(x,y) = \left( {\begin{array}{*{18}{c}}
1&x&y
\end{array}} \right){\left( {\begin{array}{*{17}{c}}
1&{{x_1^i}}&{{y_1^i}}\\
1&{{x_2^i}}&{{y_2^i}}\\
1&{{x_3^i}}&{{y_3^i}}
\end{array}} \right)^{-1}}\left( {\begin{array}{*{17}{c}}
{{u_1^i}}\\
{{u_2^i}}\\
{{u_3^i}}
\end{array}} \right)$, where  ${u^i_l}$ is the dependent variable at the $l$th vertex in the $i$th triangle ($l$ can also be considered as the local node number), $l\in (1,2,3)$ and $i\in(1,2,3,4)$. Accordingly, the 3 auxiliary variables inside one triangle can be given by $E_i(\bm r_p) =\alpha _{(i,{\bf r}_p)}^{l} u_l^i$, where $\bm r_p$ represents the coordinate of the $p$th quadrature point, and $\alpha _{(i,{\bf r}_p)}^{l}$ denotes the $l$th interpolating function corresponding to  $u^i_l$, and $l\in(1,2,3)$. We follow  Einstein summation convention for the indices of $l$, which are repeated both in the superscript of $\alpha$ and subscript of $u$.

Secondly, we construct the discretized equation of  \Eq{discre_VolumeGreen2D}  for the auxiliary variables, i.e., field value at three quadrature points  inside a single triangle as follows, 
\beq
\label{single_triangle}
\left( {\begin{array}{*{20}{c}}
{\alpha _{(i,\bm r_1)}^lu_l^i}\\
{\alpha _{(i,\bm r_2)}^lu_l^i}\\
{\alpha _{(i,\bm r_3)}^lu_l^i}
\end{array}} \right) = \left( {\begin{array}{*{20}{c}}
{E_{0}^{(i,\bm r_1)}}\\
{E_{0}^{(i,\bm r_2)}}\\
{E_{0}^{(i,\bm r_3)}}
\end{array}} \right) + k_0^2\Delta\varepsilon \frac{{{A_i}}}{3}\left( {\begin{array}{*{20}{c}}
{g_{(i,\bm r_1)}^{(i,\bm r_p)}\alpha_{(i,\bm r_p)}^l u_l^i}\\
{g_{(i,\bm r_2)}^{(i,\bm r_p)}\alpha_{(i,\bm r_p)}^l u_l^i}\\
{g_{(i,\bm r_2)}^{(i,\bm r_p)}\alpha_{(i,\bm r_p)}^l u_l^i}
\end{array}} \right) +  \sum\limits_{j\neq i} k_0^2\Delta\varepsilon \frac{{{A_j}}}{3}\left( {\begin{array}{*{20}{c}}
{g_{(i,\bm r_1)}^{(j,\bm r_p)}\alpha_{(j,\bm r_p)}^l u_l^j}\\
{g_{(i,\bm r_2)}^{(j,\bm r_p)}\alpha_{(j,\bm r_p)}^l u_l^j}\\
{g_{(i,\bm r_3)}^{(j,\bm r_p)}\alpha_{(j,\bm r_p)}^l u_l^j}
\end{array}} \right),
\eeq
where Einstein summation convention is applied for the indices of $l$ and $\bm r_p$, and $(i, \bm r_p)$ denotes the quadrature point ($\bm r_p \in \left(\bm r_1, \bm r_2, \bm r_3 \right)$) in the $i$th triangle, the label of interpolating function $l \in \left( {1,2,3} \right)$, $i/j$ indexes the triangle and does not follow  Einstein summation rule. In \Eq{single_triangle}, the 3 auxiliary variables in triangle $i$ on  the left hand side equal to the sum of three terms from the right hand side: the first term is the incident field, the second term is the self-term contribution from the same triangle $i$, the last term is the sum of the contribution from all the other triangles. $A_j$ is the area of the triangle $j$. Applying \Eq{single_triangle} to the simple scatter shown in \Fig{fig1} (c) for the 4 triangles one by one, one shall obtain the following discretized GIE in the matrix form as follows,
\beq\label{GIE_matrix}
\left( {\begin{array}{*{20}{c}}
\bar{\bm\alpha} _{11}&\bm 0&\bm 0&\bm 0\\
\bm 0&\bar{\bm\alpha} _{22}&\bm 0&\bm 0\\
\bm 0&\bm 0&\bar{\bm\alpha} _{33}&\bm 0\\
\bm 0&\bm 0&\bm 0&\bar{\bm\alpha} _{44}
\end{array}} \right)\left( {\begin{array}{*{20}{c}}
\bm U_1\\
\bm U_2\\
\bm U_3\\
\bm U_4
\end{array}} \right){\rm{ = }}\left( {\begin{array}{*{20}{c}}
\bm E_{01}\\
\bm E_{02}\\
\bm E_{03}\\
\bm E_{04}
\end{array}} \right){\rm{ + }}\left( {\begin{array}{*{20}{c}}
\bm{\bar g}_{11}&\bm{\bar g}_{12}&\bm{\bar g}_{13}&\bm{\bar g}_{14}\\
\bm{\bar g}_{21}&\bm{\bar g}_{22}&\bm{\bar g}_{23}&\bm{\bar g}_{24}\\
\bm{\bar g}_{31}&\bm{\bar g}_{32}&\bm{\bar g}_{33}&\bm{\bar g}_{34}\\
\bm{\bar g}_{41}&\bm{\bar g}_{42}&\bm{\bar g}_{43}&\bm{\bar g}_{44}
\end{array}} \right)\left( {\begin{array}{*{20}{c}}
\bm U_1\\
\bm U_2\\
\bm U_3\\
\bm U_4
\end{array}} \right),
\eeq
where $\bm U_i=\left[ u^i_1, u^i_2, u^i_3\right]^T$ is a $3 \times 1$ column vector, i.e., the 3 dependent variables for triangle $i$, and $\bar{\bm\alpha} _{ii}$ and  $\bm{\bar g} _{ij}$ are $3 \times 3$ matrices as given by \Eq{single_triangle}, explicitly $\bar{\bm\alpha} _{ii}=\left( {\begin{array}{*{16}{c}}
a_{(i,{\bm r_1})}^1&a_{(i,{\bm r_1})}^2&{a_{(i,{\bm r_1})}^3}\\
a_{(i,{\bm r_2})}^1&a_{(i,{\bm r_2})}^2&{a_{(i,{\bm r_2})}^3}\\
a_{(i,{\bm r_3})}^1&a_{(i,{\bm r_3})}^2&{a_{(i,{\bm r_3})}^3}
\end{array}} \right)$, and $\bm{\bar g}_{ij}=  k_0^2\Delta\varepsilon \frac{{{A_j}}}{3} \sum\limits_{\bm r_l \in ({\bm r_1,\bm r_2,\bm r_3})}  \left( {\begin{array}{*{16}{c}}
g_{(i,{\bm r_1})}^{(j,{\bm r_l})} & g_{(i,{\bm r_1})}^{(j,{\bm r_l})}& g_{(i,{\bm r_1})}^{(j,{\bm r_l})}\\
g_{(i,{\bm r_2})}^{(j,{\bm r_l})} & g_{(i,{\bm r_2})}^{(j,{\bm r_l})} & {g_{(i,{\bm r_2})}^{(j,{\bm r_l})}}\\
g_{(i,{\bm r_3})}^{(j,{\bm r_l})} & g_{(i,{\bm r_3})}^{(j,{\bm r_l})} & {g_{(i,{\bm r_3})}^{(j,{\bm r_l})}}
\end{array}} \right) \left( {\begin{array}{*{16}{c}}
a_{(j,{\bm r_l})}^1&a_{(j,{r_l})}^2&{a_{(j,{\bm r_l})}^3}\\
a_{(j,{\bm r_l})}^1&a_{(j,{r_l})}^2&{a_{(j,{\bm r_l})}^3}\\
a_{(j,{\bm r_l})}^1&a_{(j,{r_l})}^2&{a_{(j,{\bm r_l})}^3}
\end{array}} \right) $.

\begin{table}\centering
\caption{\label{tab1}Assembling transformation}
\resizebox{100mm}{18mm}{
\begin{tabular}{c|c}
\hline
\hline
 Triangulation matrix   &  Assembling transformation matrix \\ \hline
$\bm{\bar {E}}_{tri} = \left( {\begin{array}{*{18}{c}}
1&6&2\\
2&6&3\\
6&5&3\\
5&4&3
\end{array}} \right)$  &  $\bm{\bar T}_{asm}= \left( {\begin{array}{*{18}{c}}
1&0&0&0&0&0&0&0&0&0&0&0\\
0&0&1&1&0&0&0&0&0&0&0&0\\
0&0&0&0&0&1&0&0&1&0&0&1\\
0&0&0&0&0&0&0&0&0&0&1&0\\
0&0&0&0&0&0&0&1&0&1&0&0\\
0&1&0&0&1&0&1&0&0&0&0&0
\end{array}} \right)$  \\
\hline
\hline
\end{tabular}}
\end{table}

In the last, the matrix form of GIE in \Eq{GIE_matrix} is obtained simply by stacking the  triangles one by one, thus the shared vertices among the neighbour triangles are not taken into account. Therefore, $\bm U=[\bm U_1,\bm U_2,\bm U_3,\bm U_4]^T$ is a $12\times 1$ column vector, in contrast to 6 DOFs in total shown in \Fig{fig1} (c). By introducing a global indexing of dependent variables ${\bm U}^{g}$ as well as the assembling transformation matrix $\bm{\bar T}_{asm}$, i.e, $\bm U^{g}= \bm{\bar T}_{asm}\bm U$, the \Eq{GIE_matrix} denoted  as $\left[\bm{\bar \alpha}_{ii}\right]\bm U=\bm E^{inc}+\left[\bm{\bar g}_{ij}\right] \bm U$, can be reformulated into the final matrix form that can be readily solved using the standard linear algebra,
\beq\label{GIE_matrix_final}
\left(\bm{\bar T}_{asm} \left[\bm{\bar \alpha}_{ii}\right] \bm{\bar T}_{asm}^T\right){\bm U}^{g}=\bm{\bar T}_{asm}{\bm E}^{inc}+ \left(\bm{\bar T}_{asm}\left[\bm{\bar g}_{ij}\right] \bm{\bar T}_{asm}^T \right) {\bm U}^{g},
\eeq
where $\left[\bm{\bar \alpha}_{ii}\right]$ ($\left[\bm{\bar g}_{ij}\right]$) is the full matrix of the block matrix $\bm{\bar \alpha}_{ii}$ ( $\bm{\bar g}_{ij}$). The assembling transformation matrix  $\bm{\bar T}_{asm}$ can be obtained from the  triangulation matrix, as tabulated in \Table{tab1}. The triangulation matrix indicates how the vertices (the global node number) are connected to the local node number in each triangle. The column number of $\bm{\bar T}_{asm}$ equals to the total number of elements of the triangulation matrix ${\bm E}_{tri}$, while  the row number equals to the total number of vertices. By flattening  ${\bm E}_{tri}$ row by row, one can construct  $\bm {\bar T}_{asm}$ by mapping the row to the element of the flattened ${\bm E}_{tri}$, with the location of  nonzero element  in the same row corresponding to the value of that element in the flattened ${\bm E}_{tri}$.


\begin{figure}[ht!]
\centering
\includegraphics[scale=0.3]{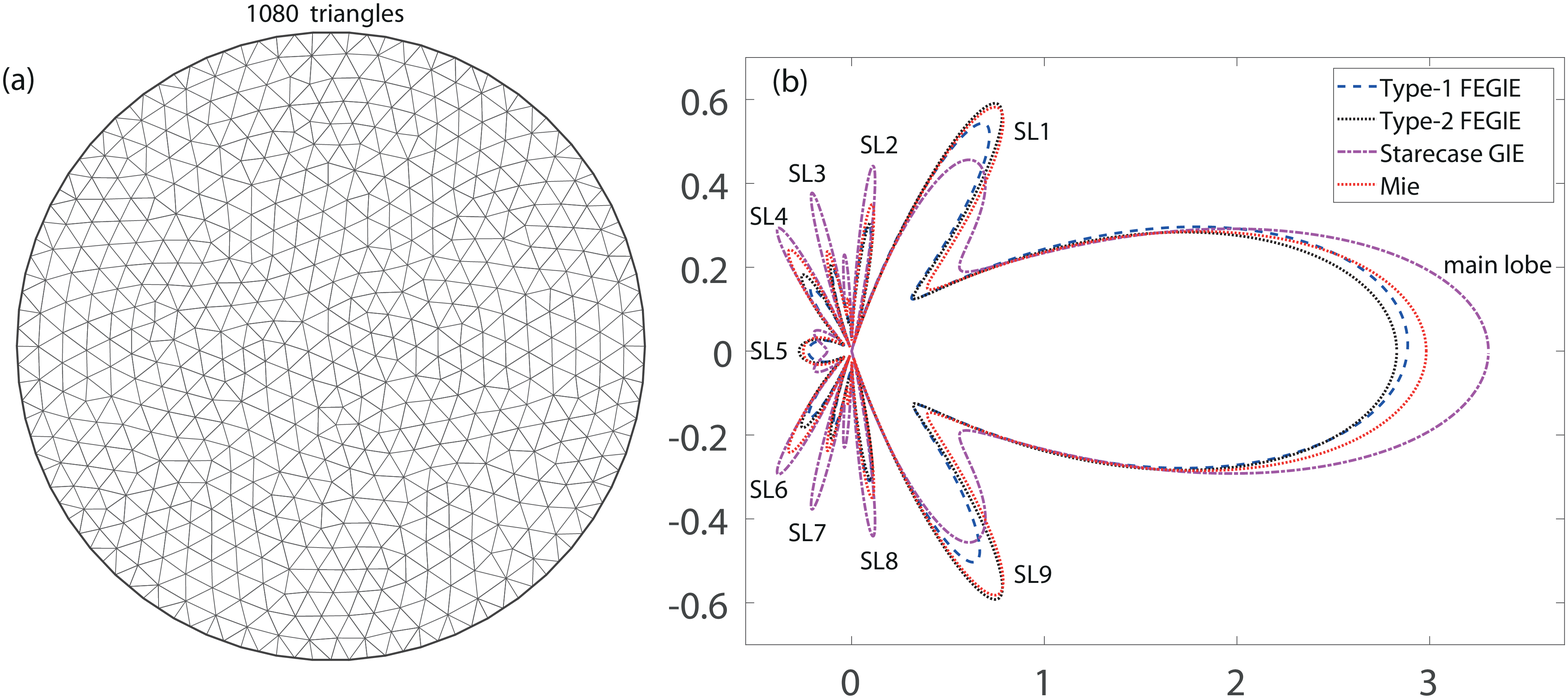}
\caption{(a) Triangle discretization of the dielectric rod. (b) Far-field pattern calculated using type-1 FEGIE, type-2 FEGIE, staircase GIE, and Mie theory. The radius of the scatter is 0.6 $\mu$m, and the vacuum wavelength is 0.633 $\mu$m. The numbers of DOFs in type-1 FEGIE, type-2 FEGIE, and staircase are 1080, 595, and 1085, respectively.}
\label{fig2}
\end{figure}

\section{Results and discussions}
We examine the far-field pattern of 2D light scattering \cite{Bohren} using our proposed method, i.e., type-1 FEGIE and type-2 FEGIE, in comparison with stair-case GIE \cite{Sondergaard2018book,sondergaard2007modeling} and Mie theory. In the first example, we study light scattering by a dielectric rod (infinitely long along z-axis, $\epsilon_r$=4), with the background being air. The incident  light is TE polarized, propagating along x-axis with vacuum wavelength 0.633 $\mu$m. Without explicit mention, the same incident conditions apply to the  scattering problems throughout this section. The dielectric rod is discretized with triangle elements shown in Fig. 2(a), and the farfield pattern calculated using Mie theory (red dotted line), type-1 FEGIE (blue dashed line), type-2 FEGIE (black dotted line), staircase FIE (magenta dash-dot line). As for the scatter with simple geometric shape, our type-1 FEGIE and type-2 FEGIE  yield reasonably accurate far-field calculations, as benchmarked against the results obtained from Mie theory as well as staircase GIE. Due to improved geometric approximation of the scatter, it is expected our proposed FEGIE is more accurate than that of staircase GIE. Indeed, the  farfield pattern calculated  from our type-1 FEGIE and type-2 FEGIE shows more accurate results, compared with staircase GIE, where the number of triangles in  type-1 FEGIE and type-2 FEGIE are the same, i.e., 1080, and the number of squares in staircase GIE is 1085. A close examination shows that type-2 FEGIE captures the sharp features of the far-field pattern much better than that of type-1 FEGIE  and  staircase GIE, due to the fact that the finite element integration technique in each element, i.e., quadrature rule, is introduced in type-2 FEGIE to improve the numerical integration. As a side remark, the numbers of DOFs in type-1 FEGIE (staircase GIE) is the same as the number of the triangles (squares), i.e. approximately 1085.  While the number of DOFs used in type-2 FEGIE is equal to the number of  vertices, i.e., 595, which is less than those in  type-1 FEGIE and staircase GIE. The smaller number of DOFs can be considered as the second advantage of type-2 FEGIE, apart from the ability of capturing the sharp features in the farfield pattern.

\begin{figure}[ht!]
\centering
\includegraphics[scale=0.3]{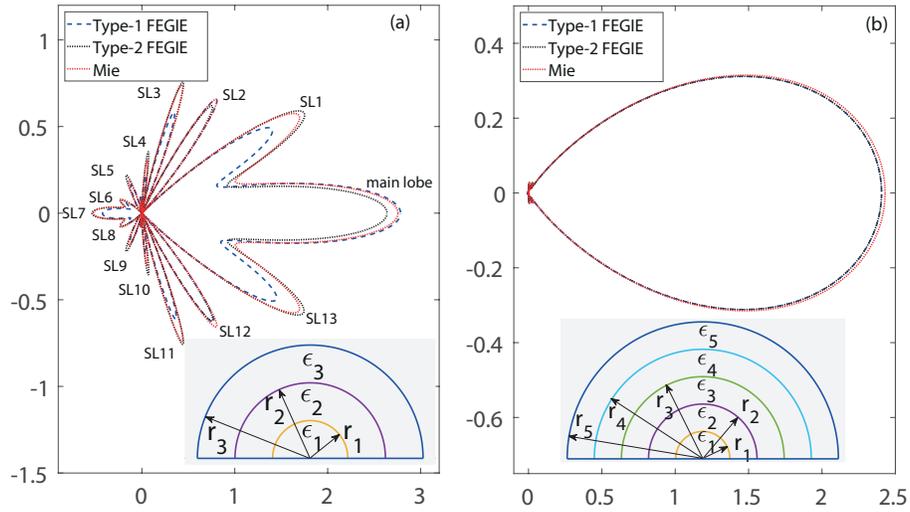}
\caption{Far-field pattern of the light scattering by (a) the 3-layered rod and (b) the 5-layered rod calculated using type-1 FEGIE, type-2 FEGIE, and COMSOL. For the 3-layered (5-layered) rod, the numbers of DOFs in type-1 FEGIE and type-2 FEGIE are 3186 (4812) and 1674 (2507).}
\label{fig3}
\end{figure}

In the second example as shown in \Fig{fig3}, we proceed to discuss light  scattering of 3-layered rod and 5-layered rod  using type-1 FEGIE (blue dashed line) and  type-2 FEGIE (black dotted line), which are again compared with finite element calculations from Mie theory (red dotted line). The staircase GIE calculation of this problem is poorly converged, which is not shown here. As shown in the inset of (a), the radii of the three layers rod are specified to be $r_1$=0.3 $\mu$m, $r_2$=0.7 $\mu$m, $r_3$=1.0 $\mu$m, and the dielectric constants of the three layers are given as $\epsilon_1=3$, $\epsilon_2= 4$, $\epsilon_3=5$. The radii of the 5-layered rod \cite{Arslanagic}, as sketched in the inset of (b),  are given by $r_1$=0.0158 $\mu$m, $r_2$=0.0285 $\mu$m, $r_3$=0.0475 $\mu$m, $r_2$=0.0601 $\mu$m, $r_5$=0.633 $\mu$m. The corresponding  dielectric constants are given as  $\varepsilon_1 =0.9814$, $\varepsilon_2=4.7676$, $\varepsilon_3=1.7075$, $\epsilon_4=1.4711$, $\varepsilon_5=1.2025$. We  examine the differences of the far-field patterns for the 3/5-layered rod calculated from type-1 FEGIE and type-2 FEGIE, as shown in \Fig{fig3} (a/b). In \Fig{fig3} (a), both type-1 FEGIE and type-2 FEGIE can capture the overall features of far-field pattern, including the angular positions and amplitudes of the main lobes and side lobes (SLs), in consistency  with calculations from Mie theory. Notably, the far-field pattern from type-2 FEGIE yields excellent agreement with that FEM calculations, except slight deviations in the main lobe and a few side lobes (SL1, SL2, SL10, SL11). In contrast, the far-field pattern from type-1 FEGIE deviates in most of the side lobes, and close examination also reveals that the far-field pattern  (blue dashed line) is not symmetric with respect with the x-axis. As a side remark, the number of DOFs used in type-1 FEGIE and type-2 FEGIE are  3186 and 1674 respectively. The light scattering of a 5-layered rod  is examined in \Fig{fig3} (b). The particular choices of the geometric parameters and dielectric functions for the 5 layers rod yields highly directional radiation pattern \cite{Arslanagic}, which can be well captured using our proposed  type-1 FEGIE and  type-2 FEGIE. Importantly, our proposed approaches yield accurate results and using much less DOFs, which is highly promising for modelling light scattering of optically large structures.

\begin{figure}[ht!]
\centering
\includegraphics[scale=0.3]{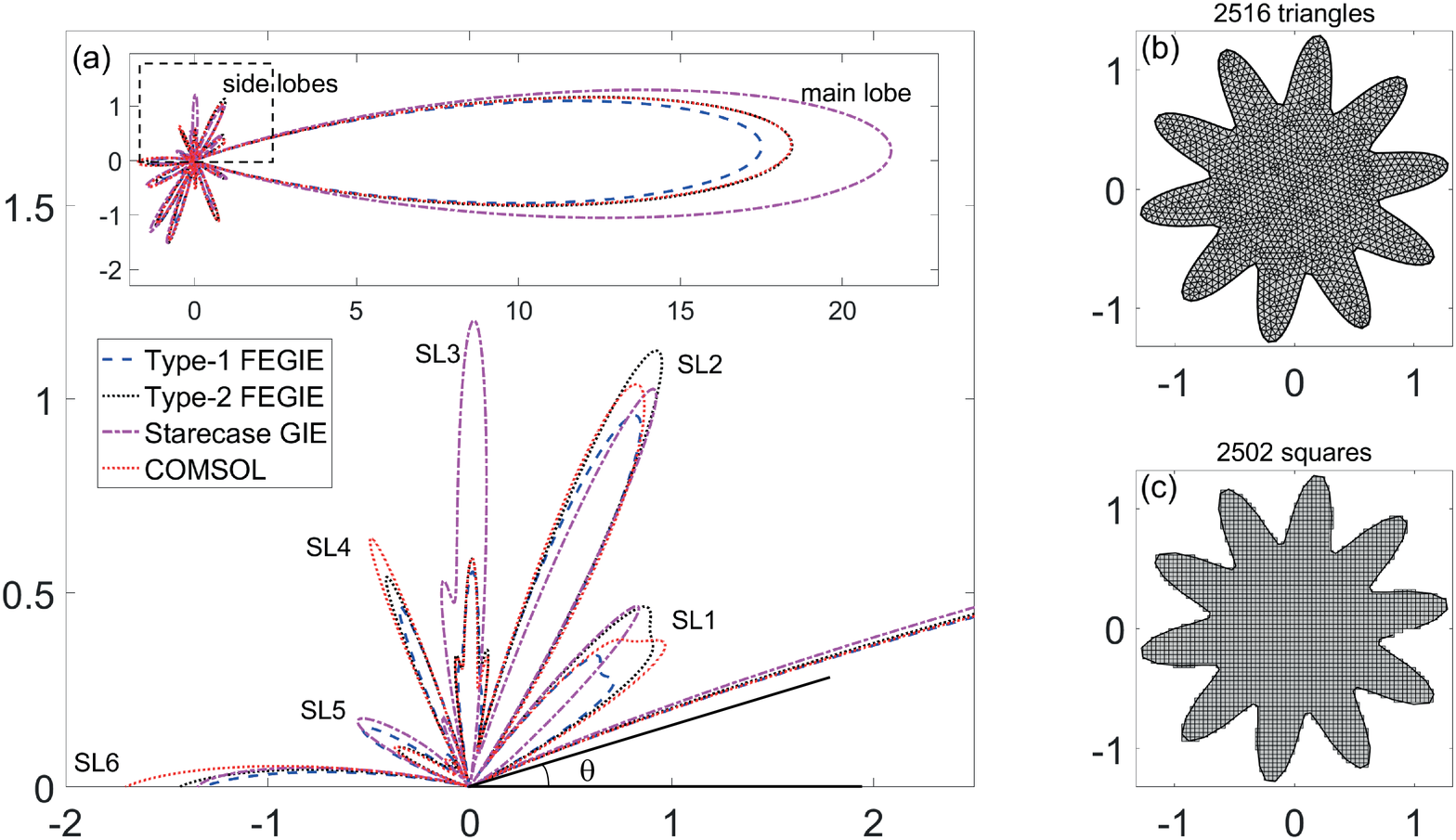}
\caption{(a) Far-field patterns of the light scattering by a complex scatter with corrugated surface calculated using type-1 FEGIE, type-2 FEGIE, staircase GIE, and Mie theory. (b/c) Triangle/Staircase approximation of the surface-corrugated scatter. The numbers of DOFs in type-1 FEGIE, type-2 FEGIE, staircase, and COMSOL are 2502, 1370, 2516, and 103561 respectively.}
\label{fig4}
\end{figure}
 
As the third example, we study the far-field pattern of a geometrically complicated scatter with corrugated surface as shown in \Fig{fig4}. The far-field pattern is calculated using type-1 FEGIE (blue dashed line), type-2 FEGIE (black dotted line), staircase FIE (magenta dash-dot line)) and COMSOL (red dotted line). The scatter with corrugated surface discretized with triangle and square elements are shown in Fig. 4(b) and (c), which have comparable number of elements, i.e., 2516 triangles versus 2502 squares. \Fig{fig4} (a) shows a few side lobes of the far-field, as indicated by the dash rectangle shown in the inset of the full far-field pattern. As can be seen from the inset and main panel of \Fig{fig4} (a), the main lobe and two side lobes (SL3 and SL5) calculated from type-2 FEGIE shows excellent agreements with that calculated from finite element method using COMSOL. In contrast, the type-1 FEGIE and staircase GIE yield considerable deviation of the main lobe and all the other side lobes. As regarding to other side lobes shown in \Fig{fig4} (a), type-2 FEGIE yet performs better in calculating far-field pattern than type-1 FEGIE and staircase GIE. It is worthy to mention that the DOFs used in type-1 FEGIE, type-2 FEGIE, staircase GIE and COMSOL are 2502, 1370, 2516 and 103561.

\begin{figure}[ht!]
\centering
\includegraphics[scale=0.4]{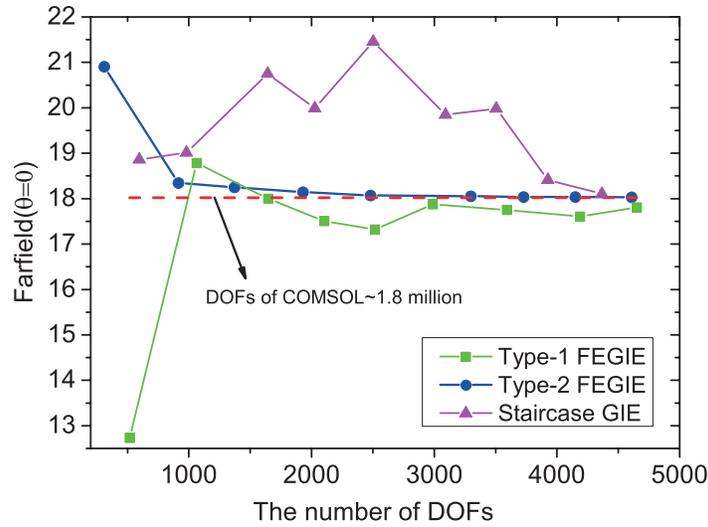}
\caption{Farfield convergence versus the number of DOFs. }
\label{fig5}
\end{figure}

Lastly, \Fig{fig5} shows the convergence of the farfield plotted in \Fig{fig4} (a) at $\theta$=0, for type-1 FEGIE (green-square), type-2 FEGIE (blue-circle), and staircase GIE (violet-triangle). The red dash line indicates the reference value of farfield obtained from COMSOL with 1.8 million DOFs. Evidently, our Type-2 FEGIE has a stable and fast convergence as the number of DOFs increases, while Type-1 FEGIE shows oscillations and a slower convergence than that of Type-2 FEGIE due to the redundant DOFs. Staircase GIE has the slowest convergence due the poor approximation of the geometry as well as the redundant DOFs. Thus, as evident from the far-field calculations and the listed DOFs used in the four approaches, our proposed type-2 FEGIE indeed shows the advantages of higher accuracy and computational efficiency due to the significant reduction of DOFs. As a final remark, the accuracy of the staircase GIE used in this paper can be improved by using averaged dielectric constant. The  averaged dielectric constant is not implemented in this paper due to the complications of calculating the differences of approximated squares with the true geometry, especially for complex structures. We also note that the singular behavior of Green's function in 2D scattering is rather weak, thus extending our approach to the 3D scattering problems with strong singular behaviors, though beyond the scope of the current paper, is certainly interesting to  investigate in the future.

\section{Conclusion}
In conclusion, we introduce finite element techniques into the Green's function integral equation, with a better and more flexible approximation of the geometry of the scatters and a more accurate numerical integral. We have calculated the farfield patterns of several concrete scatters with two types of FEGIE. We find that the two discretization schemes in FEGIE yield considerably accurate far-field pattern compared with that obtained using COMSOL, and yield more accurate far-field pattern compared with the standard discretization scheme, i.e., stair-cased approximated with square grids, which is highly promising for modelling light scattering of optically large and complex structures. Besides, the number of DOFs used in type-2 FEGIE is less than those in type-1 FEGIE and staircase GIE. 

\section*{Acknowledge}
We thank Thomas M. S\o ndergaard for fruitful discussions and a careful reading of the manuscript.
\section*{Appendix A: Scalar Green's function,  analytical approximation of self-term integration, and the far-field calculations in 2D Green's function integral equation}
The scalar Green's function of the TE polarization in 2D light scattering problems  is given by  $g\left( {{\bf{r}},{\bf{r'}}} \right) = \frac{1}{{4i}}H_0^{\left( 2 \right)}\left( {{k_0}{n_{ref}}\left| {{\bf{r}} - {\bf{r'}}} \right|} \right)$. As $i = j$ in \Eq{discre_VolumeGreen2D}, $g_{ii}$ represents the interaction of a discrete element with itself, it can be approximated with
\beq
g_{ii} = \frac{1}{{{A_i}}}\int_{{A_i}} {g\left( {{\bf{r}},{\bf{r'}}} \right)} dA'
\eeq
If we adopt square discrete elements, a square element can be approximated by circular one with the same area and same center position, Eq.(4) can be evaluated analytically resulting in

\beq\label{eq11}
{g_{ii}} \approx \frac{1}{{2i{{\left( {{k_0}{n_{ref}}a} \right)}^2}}}\left( {{k_0}{n_{ref}}aH_1^{\left( 2 \right)}\left( {{k_0}{n_{ref}}a} \right) - \frac{{2i}}{\pi }} \right)
\eeq

For far field, since $\left| {{\bf{r'}}} \right|/\left| {\bf{r}} \right| \ll 1$, the Green function can be approximated with

\beq
{g^{ff}}\left( {{\bf{r}},{\bf{r'}}} \right) = \frac{1}{4}\sqrt {\frac{2}{{\pi kr}}} {e^{\frac{{ - i\pi }}{4}}}{e^{ - ikr}}{e^{ik\frac{{\bf{r}}}{r} \cdot {\bf{r'}}}}
\eeq

Then the far field can be obtained as

\beq
E_{SC}^{ff}\left( {\bf{r}} \right) = \frac{1}{4}\sqrt {\frac{2}{{\pi kr}}} {e^{\frac{{ - i\pi }}{4}}}{e^{ - ikr}}\int {k_0^2} \left( {\varepsilon \left( {\bf{r}} \right) - {\varepsilon _{ref}}} \right)E\left( {{\bf{r'}}} \right){e^{ik\frac{{\bf{r}}}{r} \cdot {\bf{r'}}}}dA'
\eeq

\section*{Funding}
Natural National Science Foundation (NSFC) (11874026, 61735006, 61775063); National Key Research and Development Program of China (2017YFA0305200); Research Grants Council of Hong Kong SAR (CityU 21302018).

\end{document}